\documentclass[aps,prl,twocolumn]{revtex4}
\begin{document}
\draft
\def\be{\begin{equation}}
\def\ee{\end{equation}}
\def\ba{\begin{eqnarray}} 
\def\ea{\end{eqnarray}}
 
\newcommand{\bbf}{\mathbf}
\newcommand{\rrm}{\mathrm}

%\twocolumn[\hsize\textwidth\columnwidth\hsize\csname@twocolumnfalse%
%\endcsname
\title{Renormalisation and fixed points in Hilbert Space\\} 
 
\author{Tarek Khalil
%Laboratoire de Physique Th\'eorique\footnotemark[1]
and Jean Richert \\
Laboratoire de Physique Th\'eorique, UMR 7085 CNRS/ULP,\\
Universit\'e Louis Pasteur, 67084 Strasbourg Cedex,\\ 
France} 
 
\date{\today}
\begin{abstract}
The energies of low-lying bound states of a microscopic quantum many-body 
system of particles can be worked out in a reduced Hilbert space.
We present here and test a specific non-perturbative truncation procedure. 
We also show that real exceptional points which may be present in the spectrum 
can be identified as fixed points of coupling constants in the truncation 
procedure.  
\end{abstract} 
\maketitle
%%]

The ab initio construction of a rigorous quantum many-body theory able to 
describe bound particle systems like molecules, atoms, aggregates, atomic nuclei
and condensed systems has developed over a long period of time starting in the 
sixties ~\cite{brand}.  
 
In practice the explicit resolution of the problem necessitates the 
diagonalisation of the
many-body Hamiltonian in Hilbert space which is spanned by a complete set of
basis states, in principle of infinite dimension, at least generally very
large. In many cases the information of interest is however restricted to the  
knowledge of
a few energetically low-lying states which often possess collective properties.
Hence it would be convenient to work in a finite truncated subspace of the 
original Hilbert space.  
Many different procedures have been proposed and successfully applied
in the framework of the nuclear shell model and related microscopic 
descriptions.\\ 
Rigorous projection methods lead to effective Hamiltonians which
can in principle be explicitly generated by means of perturbation techniques
~\cite{brand,sch1}.
%~\cite{brand,kuo1,john,ober,sch1,sch2}.
Unfortunately there exists no straightforward control on the convergence 
properties
of the perturbation expansions which are involved, especially not when the 
interaction 
between the particles is strong as it is the case in atomic nuclei 
%~\cite{bruce}
or in quantum spin systems for instance. Many attempts have been 
made in order to overcome this problem ~\cite{suz,kum,hof,hax}. Pragmatic 
phenomenological procedures have also been introduced~\cite{ger,wil}. More
recently effective $2$-body interactions have been constructed by means of a
non-perturbative renormalisation technique which cuts-off the large momentum
components of the interaction~\cite{bog}.  

%Most of these methods do not tackle the problem related to the breakdown of 
%perturbation
%methods in specific but frequently encountered situations. This is the case in 
%the framework of the projection  method when two states belonging to different 
%subspaces come arbitrarily close to or even cross each other for reasons 
%related to
%symmetry properties~\cite{sch2}. 
 
The investigations which follow are developed in the spirit of former work
based on renormalization concepts ~\cite{gla,rau,mue,bec}.  
%~\cite{mar}. 
We introduce here a possible non-perturbative scheme for the bound state 
many-body problem which relies on a systematic reduction of Hilbert space.  
 
%We first introduce the formal framework in which we develop the 
%dimensional reduction procedure. We then define the method designed for to 
%many-body 
%quantum systems at temperature $T = 0$.  Finally we show in this framework the link 
%which exists between
%exceptional points where perturbation series expansions diverge 
%and fixed points which may characterise the renormalisation approach.

{\it Formal framework}.
Consider a
system with a fixed but arbitrary number of bound quantum objects in a Hilbert 
space $\cal H^{(N)}$ of dimension $N$ governed by a Hamiltonian
$H^{(N)}\left ( g_1^{(N)},g_2^{(N)},...g_p^{(N)}\right)$ where
$\left\{ g_1^{(N)},g_2^{(N)},\cdots, g_p^{(N)}\mapsto g^{(N)}\right\} $ 
are a set of 
parameters (coupling constants) which characterise $H^{(N)}$.
The eigenvectors $|\Psi_i^{(N)}(g^{(N)})\rangle$
$\left\{ i=1,\cdots, N\right\}$ of $H^{(N)}$
span the Hilbert space and are the solutions of the
Schr\"odinger equation
\be
H^{(N)}(g^{(N)}) |\Psi_i^{(N)}(g^{(N)})\rangle = \lambda_i(g^{(N)}) 
|\Psi_i^{(N)}(g^{(N)})\rangle  
\label{eq0} \ .
\ee
The diagonalisation of $H^{(N)}$ delivers both the eigenvalues 
$\left\{\lambda_i(g^{(N)}) , \, i=1,\cdots, N\right\}$ and
eigenvectors $\left\{|\Psi_i^{(N)}(g^{(N)})\rangle,\, i=1,\cdots,N\right\}$ 
in terms of a linear combination of orthogonal basis states
$\left\{|\Phi_i\rangle, \,i=1,\cdots, N\right\}$.
Since dim ${\cal H}^{(N)} = N$ is generally very large if not infinite and the
information needed reduces to a finite part of the spectrum it makes sense to
try to restrict the space dimensions. If the relevant
quantities of interest are for instance $M$ eigenvalues out of the set
$\left\{\lambda_i(g^{(N)})\right\}$ then one looks for a new Hamiltonian
$H^{(M)}(g^{(M)})$ such that 
\be
H^{(M)}(g^{(M)}) |\Psi_i^{(M)}(g^{(M)})\rangle = \lambda_i(g^{(M)}) 
|\Psi_i^{(M)}(g^{(M)})\rangle  
\label{eq1} \ 
\ee
with the constraints
\be
\lambda_i(g^{(M)}) = \lambda_i(g^{(N)})
\label{eq2} \ 
\ee
for $i = 1,...,M$.
Eq. (3) implies relations between the sets of coupling constants $g^{(M)}$ and 
$g^{(N)}$ 
\be
g_k ^{(M)} = f_k(g_1^{(N)},g_2^{(N)},...g_p^{(N)})
\label{eq3} \ 
\ee
with $k = 1,...,p$.
The solution of this set of equations generates the effective Hamiltonian 
$H^{(M)}$ whose spectrum in the restricted space ${\cal H}^{(M)}$ is the same 
as the corresponding  one in the original space. 
 
{\it Renormalisation algorithm for systems at temperature T = 0}.
We develop an explicit but 
general approach  which allows to implement the former procedure by
following the evolution of the
effective Hamiltonian of the system when the dimensions of the Hilbert space
are systematically reduced.   
Using the Feshbach formalism ~\cite{fesh,blo} we divide the Hilbert space
${\cal H}^{(N)}$ into two subspaces, $P{\cal H}^{(N)}$ and $Q{\cal H}^{(N)}$
with
\be
\mathrm{dim}\,  P{\cal H}^{(N)}= N-1, \quad \mathrm{dim}\, Q{\cal H}^{(N)}= 1
\label{eq10} \ .
\ee 
In the projected subspace $P{\cal H}^{(N)}$ the system with energy $E$ is 
described by the effective Hamiltonian
\be
H_{eff}(E) = P H P  +  P H Q {(E-Q H Q)}^{-1} Q H P
\label{eq11} \ . 
\ee
$H$ is written in the general form
\be
H = H_0 + g H_1 
\label{eq12} \  
\ee
where $H_0$ and $H_1$ are Hamiltonian operators and $g$ a strength parameter
(coupling constant) which takes the value $g^{(N)}$ in ${\cal H}^{(N)}$.

The complete set of basis states $\left\{|\Phi_i\rangle, \,i=1,\cdots, N\right\}$
may f.i. be chosen as the eigenvectors of $H_0$ with the corresponding 
eigenvalues 
$\left\{\epsilon_i, \,i=1,\cdots, N\right\}$.
The expression $H_{eff}(E)$ is generally the starting point of theories
which use perturbation expansions ~\cite{brand}. Here we proceed
differently. We consider 
\be
P |\Psi_1^{(N)}\rangle = \sum_{i=1}^{N-1} a_{1i}^{(N)}(g^{(N)})|\Phi_i\rangle
\label{eq13} \  
\ee
which is the projection on $P{\cal H}^{(N)}$ of an eigenvector 
\be
|\Psi_1^{(N)}\rangle = \sum_{i=1}^{N}  a_{1i}^{(N)}(g^{(N)})|\Phi_i\rangle
\label{eq14} \  
\ee
of ${\cal H}^{(N)}$. If $\lambda_1^{(N)}$ is the eigenvalue
corresponding to  $|\Psi_1^{(N)}\rangle$ we look for the solution of 
\be
H_{eff}(\lambda_1^{(N)})P |\Psi_1^{(N)}\rangle =  \lambda_1^{(N)}
P |\Psi_1^{(N)}\rangle
\label{eq15} \ . 
\ee
We consider $P |\Psi_1^{(N)}\rangle$ to be the lowest energy eigenstate and 
$Q{\cal H}^{(N)}$
to contain f. i. the highest one in energy. Any other state may in principle
be chosen.
We impose the lowest eigenvalue in the 
$P{\cal H}^{(N)}$ subspace to be the same as the one in the complete space
\be
\lambda_1^{(N-1)} = \lambda_1^{(N)} 
\label{eq16} \ . 
\ee
Projecting the expression of Eq.~(\ref{eq15}) on $\langle\Phi_1|$ which is the
eigenvector of $H_0$ with lowest energy 
\be
\langle\Phi_1|H_{eff}(\lambda_1^{(N)})|P \Psi_1^{(N)}\rangle =\lambda_1^{(N)}
(g^{(N)}) a_{1i}^{(N)}(g^{(N)})
\label{eq17} \ . 
\ee
and defining the Hamiltonian which acts in the projected space $P{\cal H}^{(N)}$
as $H^{(N-1)} = H_0 + g^{(N-1)} H_1$,  Eq.~(\ref{eq16}) leads to
\be
\langle\Phi_1|H_{eff}(\lambda_1^{(N)})|P \Psi_1^{(N)}\rangle = {\mathcal
F}(g^{(N-1)})
\label{eq18} \ . 
\ee
where
\be
{\mathcal F}(g^{(N-1)}) = \overline{H_{1N}^{(N-1)}} + H_{1N}
(\lambda_1^{(N)} -  H_{NN}^{(N-1)})^{-1}\overline{H_{N1}^{(N-1)}}
\label{eq19} \ 
\ee 
with  
\be
H_{ij}^{(N-1)} = \langle \Phi_i|H^{(N-1)}|\Phi_j\rangle 
\label{eq20} \  
\ee
and
\be 
\overline{H_{1N}^{(N-1)}} =  \langle \Phi_1|H^{(N-1)}|P \Psi_1^{(N)}\rangle.  
\label{eq21} \  
\ee
$\overline{H_{N1}^{(N-1)}}$ is the same expression as 
$\overline{H_{1N}^{(N-1)}}$
with $\langle \Phi_1|$ replaced by $\langle \Phi_N|$.  

The r.h.s. of Eq.~(\ref{eq18}) can be worked out explicitly. 
The denominator in the second term of  
$H_{eff}(\lambda_1^{(N)})$ of Eq.~(\ref{eq19})
is a scalar quantity since 
$\mathrm{dim}\, Q{\cal H}^{(N)}= 1$
where $N$ designates the highest-lying energy eigenstate of $H^0$,  
$|\Phi_N\rangle$.  
Imposing the constraint introduced through Eq.~(\ref{eq16}) leads to a 
relation  which fixes $g^{(N-1)}$.
One gets explicitly
\be
{a^{(N-1)}{g^{(N-1)}}^2 + b^{(N-1)}g^{(N-1)} +  c^{(N-1)}} = 0
\label{eq22} \  
\ee 
where 
\be
a^{(N-1)} = G_{1N} - H_{NN} F_{1N}
\label{eq23} \  
\ee
with
\be
H_{ij} = \langle \Phi_i|H_1|\Phi_j\rangle 
\label{eq24} \  
\ee 
\be
b^{(N-1)} =  a_{11}^{(N)} H_{NN}(\lambda_1^{(N)} - \epsilon_1) +  F_{1N}
(\lambda_1^{(N)} - \epsilon_N)
\label{eq25} \  
\ee
\be
c^{(N-1)} =   -a_{11}^{(N)}(\lambda_1^{(N)} - \epsilon_1)
(\lambda_1^{(N)} - \epsilon_N))
\label{eq26} \  
\ee
with
\be
F_{1N} =  \sum_{i=1}^{N-1} a_{1i}^{(N)} \langle \Phi_1|H_1| \Phi_i \rangle 
\label{eq27} \ . 
\ee
and
\be
G_{1N} =  H_{1N} \sum_{i=1}^{N-1} a_{1i}^{(N)} \langle \Phi_N|H_1| \Phi_i \rangle
\label{eq28} \ . 
\ee
Since Eq.~(\ref{eq22}) is non-linear in $g^{(N-1)}$ and has two solutions, 
$g^{(N-1)}$ is chosen as the one closest to $g^{(N)}$ by continuity. 
The process can be iterated  step by step by projection from the space of 
dimension $N-1$ to
$N-2$ and further, generating subsequently a succession of values of
the strength parameter (coupling constant)
$g^{(k)}$ at each iteration. At each step the projected wavefunction
$|P \Psi_1^{(k)}\rangle$ is obtained from $|\Psi_1^{(k)}\rangle$
by elimination of a state $|\Phi_k\rangle$. Taking the continuum limit for large 
$N$ leads to a differential flow equation for $g$.
%Going over 
%to the continuum limit if $N$ is large, one can derive a non-linear flow 
%equation
%\frac{dg}{dx} = - {\frac{1}{2a(x)g(x)+b(x)}}(\frac{dc}{dx} +
%\be
%\frac{dg}{dx} = - {\frac{1}{2a(x)g(x)+b(x)}}(\frac{dc}{dx} +
%{\frac{db}{dx}}g(x) + {\frac{da}{dx}}{g(x)^2})
%\label{eq29} \  
%\ee
%where $a(x), b(x), c(x)$ and $g(x)$ are the continuous extensions of the
%corresponding discrete quantities. Eq.~(\ref{eq29}) is a
%non-linear differential equation which a priori can only be solved numerically. 
%We apply now the procedure to an example in order to analyse and 
%discuss its practical efficiency.

{\it  Example}.

%We consider a system whose Hamiltonian $H = H_0 + g H_1$ is such that $H_0$ is
%diagonal in the basis of states $\left\{|\Phi_i\rangle, \,i=1,\cdots, N\right\}$
%with degenerate energies $\epsilon$. The matrix elements of $H_1$ are such that   
%\be
%\langle \Phi_i|H_1|\Phi_i\rangle = \beta  
%\label{eq30} \  
%\ee
%and
%\be
%\langle \Phi_i|H_1|\Phi_{i+1}\rangle = \langle \Phi_i|H_1|\Phi_{i-1}\rangle 
%=\gamma  
%\label{eq31} \  
%\ee

We consider a real symmetric tight-binding Hamiltonian with $H_0 = 0$,  
diagonal elements $\langle \Phi_i|H_1|\Phi_i\rangle = \beta$ and non-diagonal 
ones $\langle \Phi_i|H_1|\Phi_{i+1}\rangle = \langle \Phi_i|H_1|\Phi_{i-1}\rangle 
=\gamma$ which is typical of strongly coupled systems. 
By essence, a reduction of Hilbert space by means of perturbation expansions is
meaningless in this context. 
Starting with an initial Hilbert space dimension $N$ we apply the renormalisation
procedure described above to $g$ starting from an initial value $g^{(N)}$. The 
evolution of the lowest eigenvalues and the flow of $g$ are shown in Table 1 
for different values of $g$ and $N$. Several conclusions can be drawn. 
First, as expected, the stability of the spectrum of low-lying states is the better
the smaller the non-diagonal 
coupling between neighbouring states. Second, the stability 
increases when the initial dimension of the space increases. Third, the coupling 
constant may considerably change, decreasing in the present case.  
%We also worked on other strongly coupled systems which we shall present in a later 
%publication. In general terms, the procedure works well as long as states 
%which are strongly coupled to the states which contribute in an essential way  
%to the lowest eigenstates of the spectrum are taken into account\\ 
As a general prescription for the algorithm to work one should always keep the states 
which are strongly coupled those basis states which contribute in an essential way to 
the low-lying states of the physical spectrum.\\
\begin{center}
\begin{tabular}{|c|c|c|c|c|c|c|c|} \hline
N  & n  & g    & $\lambda_1$ & $\lambda_2$ & $\lambda_3$ & $\lambda_4$ & $\lambda_5$ \\ \hline\hline
10 & 10 & 20   & 0.81        & 3.17        & 6.90        & 11.69       & 17.15       \\ \hline
   & 7  & 13.4 & 1.02        & 3.93        & 8.29        & 13.43       & 18.57       \\ \hline
   & 5  & 8.18 & 1.10        & 4.09        & 8.18        & 12.27       & 15.26       \\ \hline\hline
20 & 20 & 20   & 0.22        & 0.89        & 1.98        & 3.47        & 5.34        \\ \hline
   & 10 & 3.28 & 0.26        & 1.04        & 2.26        & 3.83        & 5.62        \\ \hline
   & 5  & 1.13 & 0.30        & 1.13        & 2.25        & 3.38        & 4.21        \\ \hline\hline
30 & 30 & 20   & 0.10        & 0.41        & 0.92        & 1.62        & 2.51        \\ \hline
   & 15 & 6.02 & 0.12        & 0.46        & 1.02        & 1.77        & 2.68        \\ \hline
   & 5  & 1.03 & 0.14        & 0.52        & 1.03        & 1.55        & 1.93        \\ \hline\hline
50 & 50 & 20   & 0.04        & 0.15        & 0.34        & 0.60        & 0.94        \\ \hline
   & 20 & 3.73 & 0.04        & 0.17        & 0.37        & 0.65        & 0.99        \\ \hline
   & 5  & 0.38 & 0.05        & 0.19        & 0.38        & 0.57        & 0.71        \\ \hline\hline
20 & 20 & 1.0  & 0.01        & 0.04        & 0.10        & 0.17        & 0.27        \\ \hline
   & 10 & 0.33 & 0.01        & 0.05        & 0.11        & 0.19        & 0.28        \\ \hline
   & 5  & 0.11 & 0.015       & 0.06        & 0.11        & 0.17        & 0.21        \\ \hline
\end{tabular}\\
\end{center}
\begin{center}
{\it  Table 1.}\\
\end{center} 
Evolution of the coupling constant and the 5 lowest eigenvalues of the
tight-binding matrix described in the text. Here 
%$\epsilon = 0$, 
$\beta = 1$,
$\gamma = 0.5$. $N$ is the initial space dimension, $n$ the restricted dimension
and $g$ the running coupling constant.\\

{\it Exceptional points and fixed points}.\\ 
In the present renormalisation scheme exceptional points which generate divergences in 
perturbation expansions are related to the existence of fixed points of the coupling 
constant $g$. 

It has been rigorously established that the eigenvalues $\lambda_k{(g)}$
of $H(g) = H_0 + gH_1$ are analytic functions of $g$ with only algebraic
singularities~\cite{kat,hei,sch2}. They get singular at so called exceptional points 
$g = g_e$ which are first order branch points in the complex $g$ - plane. 
Branch points appear if two (or more) eigenvalues get degenerate. This can 
happen if $g$ takes values such that $H_{kk} =  H_{ll}$ 
where $H_{kk} = \langle \Phi_k|H|\Phi_k\rangle$ 
which corresponds to a so-called level crossing. As a consequence, if a level
belonging to the $P{\cal H}$ subspace defined above crosses a level
lying in the complementary $Q{\cal H}$ subspace the perturbation development
constructed from $H_{eff}(E)$ diverges~\cite{sch2}. Exceptional points are 
defined as the solutions of ~\cite{hei}
\be
f(\lambda(g_e))  = det[ H(g_e) - \lambda(g_e)I] = 0
\label{eq36} \ 
\ee
and 
\be
\frac{df(\lambda(g_e))}{d\lambda}|_{\lambda= \lambda(g_e)}= 0
\label{eq37} \ 
\ee
where $f(\lambda(g))$ is the secular determinant.
It is now possible to show that exceptional points are
connected to fixed points corresponding to $dg/dx = 0$ in specific cases. If   
$\left\{\lambda_i(g)\right\}$ are the set of eigenvalues the secular equation
can be written as
\be
\prod_{i=1}^N {(\lambda - \lambda_i)} = 0
\label{eq38} \ .
\ee
Consider $\lambda = \lambda_p$ which satisfies Eq.~(\ref{eq36}). Then
Eq.~(\ref{eq37}) can only be satisfied if there exists another eigenvalue
$\lambda_q = \lambda_ p$, hence if a degeneracy appears in the spectrum. 
This is the case at an exceptional point. 

Going back to the algorithm described above consider the case where the 
fixed
eigenvalue $\lambda_1$ gets degenerate with some other eigenvalue 
$\lambda_i^{(k)}({g = g_e})$ at some step $k$ in the space reduction process. 
Since $\lambda_1$ is constrained to be constant,
\be
\lambda_i^{(k)}({g_e}) = \lambda_i^{(l)}({g'_e})
\label{eq39} \ 
\ee
which is realised in any projected subspace of size $k$ and $l$ containing  
states $|\Phi_1\rangle$ and $|\Phi_i\rangle$. Going over to the continuum 
limit for large values of $N$ and considering the subspaces of dimension $x$ 
and $x + dx$ in this limit one can write 
\be
\frac{d\lambda_1}{dx} = 0 = \frac{d\lambda_i(x)}{dx}
\label{eq40} \ .
\ee
Consequently
\be
\frac{d\lambda_i}{dg_e} \frac{dg_e}{dx} = 0
\label{eq41} \ .
\ee
Due to the Wigner - Neumann avoided crossing rule the degeneracy of eigenvalues 
is generally not fulfilled  for real values of the coupling constant and the 
derivative of $\lambda_i$ with respect to $g$ vanishes. 
There exist however specific situations, like systems with special symmetry 
properties 
%~\cite{sch2,he,yu} 
~\cite{sch2,yu} 
or infinite systems~\cite{sml} for which 
degeneracy for real $g$ can occur. In these cases Eq.~(\ref{eq41}) is realised if
\be
\frac{dg_e}{dx} = 0\,\,\, \quad {\mathrm{and}}\,\,\,\, 
\frac{d\lambda_i}{dg_e}\not = 0
\label{eq42} \ 
\ee  
The second relation works if crossing takes place 
and $g_e$ is a fixed point in the sense of renormalisation theory.

Eq.~(\ref{eq42}) shows the connection between exceptional and fixed points 
in the framework of the present approach. Ground state degeneracy due to level 
crossing is indeed 
a signature for the existence of phase transitions~\cite{sach1}, perturbation 
expansions break down at these points. The ground state wavefunction changes its 
properties when 
the (real) coupling constant $g$ crosses the exceptional point $g_e$. There
the eigenstates exchange the main components of their projection on the set of 
basis states ${|\Phi_i \rangle, i = 1,...N}$. 
It is worthwhile to notice that the same result is not restricted to the ground state,
it is valid at any physical level crossing in which one of the eigenvalues
stays constant for any value of the coupling constant. We have tested this property on 
several systems.  
 
{\it  Example}.

Consider an $N \times N$ matrix which corresponds to a highly degenerate and strongly 
coupled system located at a fixed point whose representative matrix possesses equal
diagonal matrix elements ($H_0 = 0, H_{ii} = -0.5$ in the present example) and equal non-diagonal
elements ($H_{ij} = +0.5$). $H_{ij}$ is defined in Eq.~(\ref{eq24}). 
The diagonalisation and the application of the Hilbert space
reduction procedure leads to rigorously stable lowest eigenstates equal to $-20$ 
and the coupling constant which characterises the system stays at its initial value, here 
$g=20$, for any value of $N$. The system is located at a fixed point corresponding to a state level 
crossing (exceptional point) as discussed above. 

%We consider a quantum spin $1/2$ ladder with 4 sites $(1, 2, 1', 2')$ with
%$(1,2)$ and $(1',2')$
%located on parallel lines, on a ladder with two rungs $(1,1')$, $(2,2')$. Their 
%antiferromagnetic interaction is described by the Hamiltonian~\cite{lin}
%\be
%H = 2J_1(S_1*S^{'}_1 + S_2*S^{'}_2 +  \alpha_2(S_1*S_2 + S^{'}_1*S^{'}_2) + 
%\alpha_3S_1*S^{'}_2 + \alpha_4S^{'}_1*S_2)
%\label{eq43} \
%\ee
%where $J_1$ is a positive coupling constant which will be renormalised through the
%Hilbert space reduction procedure, $\alpha_2$, $\alpha_3$ and 
%$\alpha_4$ positive constant quantities. In the present case $H_0=0$
%and the basis vectors are chosen as $| m_1,m_1',m_2,m_2'\rangle$ 
%where $m_i = +(1/2)$ or $-(1/2)$ is the projection of the spin $1/2$ on the 
%quantification axis at site $i$.
%The subspace corresponding to $M_{tot} = 0$ where $M_{tot}$ is the sum of the spin
%projections contains 6 states. The diagonalisation for fixed $J_1$ shows that the
%eigenstates cut each other at specific values of the $\alpha_i$'s, see Fig. 1. One
%checks on this example that $J_1$ as well as the lowest eigenvalues of the spectrum
%remain fixed at the crossing points which are exceptional points.

{\it Conclusions.}\\
In summary, we developed a non-perturbative effective theory of the bound
state many-body quantum problem based on a reduction process of the dimensions
of the initial Hilbert space. The central point concerns the renormalisation of
the coupling constant which characterises the initial Hamiltonian under
the constraint that the lowest eigenenergy which corresponds to the ground state
of the system should not change. 
%We presented a projection approach for the case of a 
%system at temperature $T=0$ and constructed the flow equation for the coupling 
%constant. 
%Finally 
We showed the relationship which exists between exceptional
points corresponding to level crossings in the spectrum where perturbation expansions
break down and fixed points of the coupling constants which characterise phase
transitions. The approach was tested and discussed on explicit examples of strongly
coupled and highly degenerate systems encountered in condensed matter physics.
The formalism can be applied to other physical quantum systems like atoms, molecules, 
aggregates as well. It can be extended to systems characterised by 
several coupling constants and at finite temperature. Effective operators acting in 
reduced space can be worked out. We shall present these developments 
in forthcoming work.

%The study of the structure of other systems like atoms or nuclei can certainly be 
%performed in the same framework and may even be more easy to implement in these
%cases for reasons related to the structure of these microscopic quantum systems.\\  
%We shall show in forthcoming work that the present approach can also be 
%straightforwardly extended to the case of an Hamiltonian characterised by several 
%coupling constants and to the case of systems at finite temperature. Effective 
%expressions of observables other than the energy and flow equations for the physical
%constants which characterise them can be worked out. Other renormalisation schemes
%can be considered and will be studied in the near future.\\   

One of us (J.R.) would like to thank J. Polonyi, H. A. Weidenmueller, J. M. Carmona,
M. Henkel, D. W. Heiss and I. Rotter for their encouragements, comments, critics 
and advices.

\end{document}